\documentclass[preprint,aps,prb,superscriptaddress]{revtex4-1}
\usepackage{graphicx}
\usepackage[caption=false]{subfig}
\usepackage[version=4]{mhchem}
\usepackage{mathtools}
\usepackage{physics}
\usepackage{amsmath}
\usepackage{siunitx}
\usepackage{gensymb}
\usepackage{textcomp}
\usepackage{newtxmath}
\usepackage{multirow}
\usepackage{makecell}
\usepackage{booktabs}
\usepackage{xcolor}
\usepackage[utf8]{inputenc}
\usepackage{lineno}

\draft
\begin{document}

\title{Room-Temperature Intrinsic and Extrinsic Damping in Polycrystalline Fe Thin Films}
\author{Shuang Wu}
\affiliation{Department of Physics, Virginia Polytechnic Institute and State University, Blacksburg, VA 24061, USA}
\author{David A. Smith}
\affiliation{Department of Physics, Virginia Polytechnic Institute and State University, Blacksburg, VA 24061, USA}

\author{Prabandha Nakarmi}
\affiliation{Department of Physics and Astronomy, The University of Alabama, Tuscaloosa, AL 35487 USA}

\author{Anish Rai}
\affiliation{Department of Physics and Astronomy, The University of Alabama, Tuscaloosa, AL 35487 USA}

\author{Michael Clavel}
\affiliation{Department of Electrical and Computer Engineering, Virginia Polytechnic Institute and State University, Blacksburg, VA 24061, USA}
\author{Mantu K. Hudait}
\affiliation{Department of Electrical and Computer Engineering, Virginia Polytechnic Institute and State University, Blacksburg, VA 24061, USA}

\author{Jing Zhao}
\affiliation{Department of Geosciences, Virginia Polytechnic Institute and State University, Blacksburg, VA 24061, USA}

\author{F. Marc Michel}
\affiliation{Department of Geosciences, Virginia Polytechnic Institute and State University, Blacksburg, VA 24061, USA}

\author{Claudia Mewes}
\affiliation{Department of Physics and Astronomy, The University of Alabama, Tuscaloosa, AL 35487 USA}

\author{Tim Mewes}
\affiliation{Department of Physics and Astronomy, The University of Alabama, Tuscaloosa, AL 35487 USA}
\author{Satoru Emori}
\affiliation{Department of Physics, Virginia Polytechnic Institute and State University, Blacksburg, VA 24061, USA}

\begin{abstract}
	We examine room-temperature magnetic relaxation in polycrystalline Fe films. Out-of-plane ferromagnetic resonance (FMR) measurements reveal Gilbert damping parameters of $\approx$ 0.0024 for Fe films with thicknesses of 4-25 nm, regardless of their microstructural properties. The remarkable invariance with film microstructure strongly suggests that intrinsic Gilbert damping in polycrystalline metals at room temperature is a local property of nanoscale crystal grains, with limited impact from grain boundaries and film roughness. By contrast, the in-plane FMR linewidths of the Fe films exhibit distinct nonlinear frequency dependences, indicating the presence of strong extrinsic damping. To fit our in-plane FMR data, we have used a grain-to-grain two-magnon scattering model with two types of correlation functions aimed at describing the spatial distribution of inhomogeneities in the film. However, neither of the two correlation functions is able to reproduce the experimental data quantitatively with physically reasonable parameters. Our findings advance the fundamental understanding of intrinsic Gilbert damping in structurally disordered films, while demonstrating the need for a deeper examination of how microstructural disorder governs extrinsic damping.
\end{abstract}
\maketitle
\section{Introduction}
In all magnetic materials, magnetization has the tendency to relax toward an effective magnetic field. How fast the magnetization relaxes governs the performance of a variety of magnetic devices. For example, magnetization relaxation hinders efficient precessional dynamics and should be minimized in devices such as precessional magnetic random access memories, spin-torque oscillators, and magnonic circuits\cite{Diao2007,Zhu2007,Yu2014,Rowlands2019}. From the technological perspective, it is important to understand the mechanisms behind magnetic relaxation in thin-film materials that comprise various nanomagnetic device applications. Among these materials, bcc Fe is a prototypical elemental ferromagnet with attractive properties, including high saturation magnetization, soft magnetism\cite{kanada2017soft}, and large tunnel magnetoresistance\cite{Parkin2004, Ando2005}. Our present study is therefore motivated by the need to uncover magnetic relaxation mechanisms in Fe thin films -- particularly polycrystalline films that can be easily grown on arbitrary substrates for diverse applications.

To gain insights into the contributions to magnetic relaxation, a common approach is to examine the frequency dependence of the ferromagnetic resonance (FMR) linewidth. The most often studied contribution is viscous Gilbert damping\cite{gilbert1955,Gilbert2004,Heinrich1967,Kambersky1976,Tserkovnyak2004,Rossi2005}, which yields a linear increase in FMR linewidth with increasing precessional frequency. In ferromagnetic metals, Gilbert damping arises predominately from ``intrinsic" mechanisms~\cite{Schoen2016,Gilmore2007,Mankovsky2013} governed by the electronic band structure\footnote{Eddy-current damping \cite{Scheck2006} and radiative damping \cite{schoen2015radiative} can also contribute to viscous damping, but they typically constitute a small correction that is $\lesssim$10\% of intrinsic Gilbert damping in ferromagnetic thin films (i.e., $\lesssim$20 nm thick) for nanomagnetic devices~\cite{Khodadadi2020, Smith2020}, which are thought to be rooted in the electronic band structure of the ferromagnetic metal~\cite{Schoen2016,Gilmore2007,Mankovsky2013}}.  Indeed, a recent experimental study by Khodadadi \textit{et al.}\cite{Khodadadi2020} has shown that intrinsic, band-structure-based Gilbert damping dominates magnetic relaxation in high-quality crystalline thin films of Fe, epitaxially grown on lattice-matched substrates. However, it is yet unclear how intrinsic damping is impacted by the microstructure of polycrystalline Fe films. 

Microstructural disorder in polycrystalline Fe films can also introduce \emph{extrinsic} magnetic relaxation. A well-known extrinsic relaxation mechanism is two-magnon scattering, where the uniform precession mode with zero wave vector scatters into a degenerate magnon mode with a finite wave vector~\cite{Geschwind1957,LeCraw1958,Schloemann1958a, Patton1967}. Two-magnon scattering generally leads to a nonlinear frequency dependence of the FMR linewidth, governed by the nature of magnon scattering centers at the surfaces \cite{Arias1999,Arias2000} or in the bulk of the film\cite{McMichael2004, woltersdorf2004two, Mo2005,Kalarickal2009}. While some prior experiments point to the prominent roles of extrinsic magnetic relaxation in polycrystalline ferromagnetic films~\cite{Lindner2009,Jiang2017, Edwards2019},  systematic studies of extrinsic relaxation (e.g., two-magnon scattering) on polycrystalline Fe thin films are still lacking. 

Here, we investigate both the intrinsic and extrinsic contributions to magnetic relaxation at room temperature in polycrystalline Fe films. 
We have measured the frequency dependence of the FMR linewidth with (1) the film magnetized out-of-plane (OOP), where two-magnon scattering is suppressed~\cite{McMichael2004} such that intrinsic Gilbert damping is quantified reliably, and (2) the film magnetized in-plane (IP), where two-magnon scattering is generally expected to coexist with intrinsic Gilbert damping. 

From OOP FMR results, we find that the intrinsic Gilbert damping of polycrystalline Fe films at room temperature is independent of their structural properties and almost identical to that of epitaxial films. Such insensitivity to microstructure is in contrast to disorder-sensitive Gilbert damping recently shown in epitaxial Fe at \emph{cryogenic} temperature~\cite{Khodadadi2020}. Our present work implies that Gilbert damping at a sufficiently high temperature becomes a local property of the metal, primarily governed by the structure \emph{within} nanoscale crystal grains rather than grain boundaries or interfacial disorder. This implication refutes the intuitive expectation that intrinsic Gilbert damping should depend on structural disorder in polycrystalline films. 

In IP FMR results, the frequency dependence of the FMR linewidth exhibits strong nonlinear trends that vary significantly with film microstructure. To analyze the nonlinear trends, we have employed the grain-to-grain two-magnon scattering model developed by McMichael and Krivosik\cite{McMichael2004} with two types of correlation functions for capturing inhomogeneities in the film. However, neither of the correlation functions yields quantitative agreement with the experimental results or physically consistent, reasonable parameters. This finding implies that a physical, quantitative understanding of extrinsic magnetic relaxation requires further corrections of the existing two-magnon scattering model, along with much more detailed characterization of the nanoscale inhomogeneities of the magnetic film. Our study stimulates opportunities for a deeper examination of fundamental magnetic relaxation mechanisms in structurally disordered ferromagnetic metal films. 

\section{Film deposition and structural properties}
\label{sec:structure}
Polycrystalline Fe thin films were deposited using DC magnetron sputtering at room temperature on Si substrates with a native oxide layer of SiO$_2$. The base pressure of the chamber was below $1\times10^{-7}$ Torr and all films were deposited with 3 mTorr Ar pressure. Two sample series with different seed layers were prepared in our study: subs./Ti(3 nm)/Cu(3 nm)/Fe(2-25 nm)/Ti(3 nm) and subs./Ti(3 nm)/Ag(3 nm)/Fe(2-25 nm)/Ti(3 nm). In this paper we refer to these two sample series as Cu/Fe and Ag/Fe, respectively. The layer thicknesses are based on deposition rates derived from x-ray reflectivity (XRR) of thick calibration films. The Ti layer grown directly on the substrate ensures good adhesion of the film, whereas the Cu and Ag layers yield distinct microstructural properties for Fe as described below. We note that Cu is often used as a seed layer for growing textured polycrystalline ferromagnetic metal films~\cite{Ghosh2012,Schoen2017}. Our initial motivation for selecting Ag as an alternative seed layer was that it might promote qualitatively different Fe film growth~\cite{BurglerFeOnAg1997}, owing to a better match in bulk lattice parameter $a$ between Fe ($a \approx 2.86$ \AA) and Ag ($a/\sqrt{2} \approx 2.88$ \AA) compared to Fe and Cu ($a/\sqrt{2} \approx 2.55$ \AA).

\begin{figure}[!htbp]
	\centering
	\includegraphics[width=0.31\linewidth]{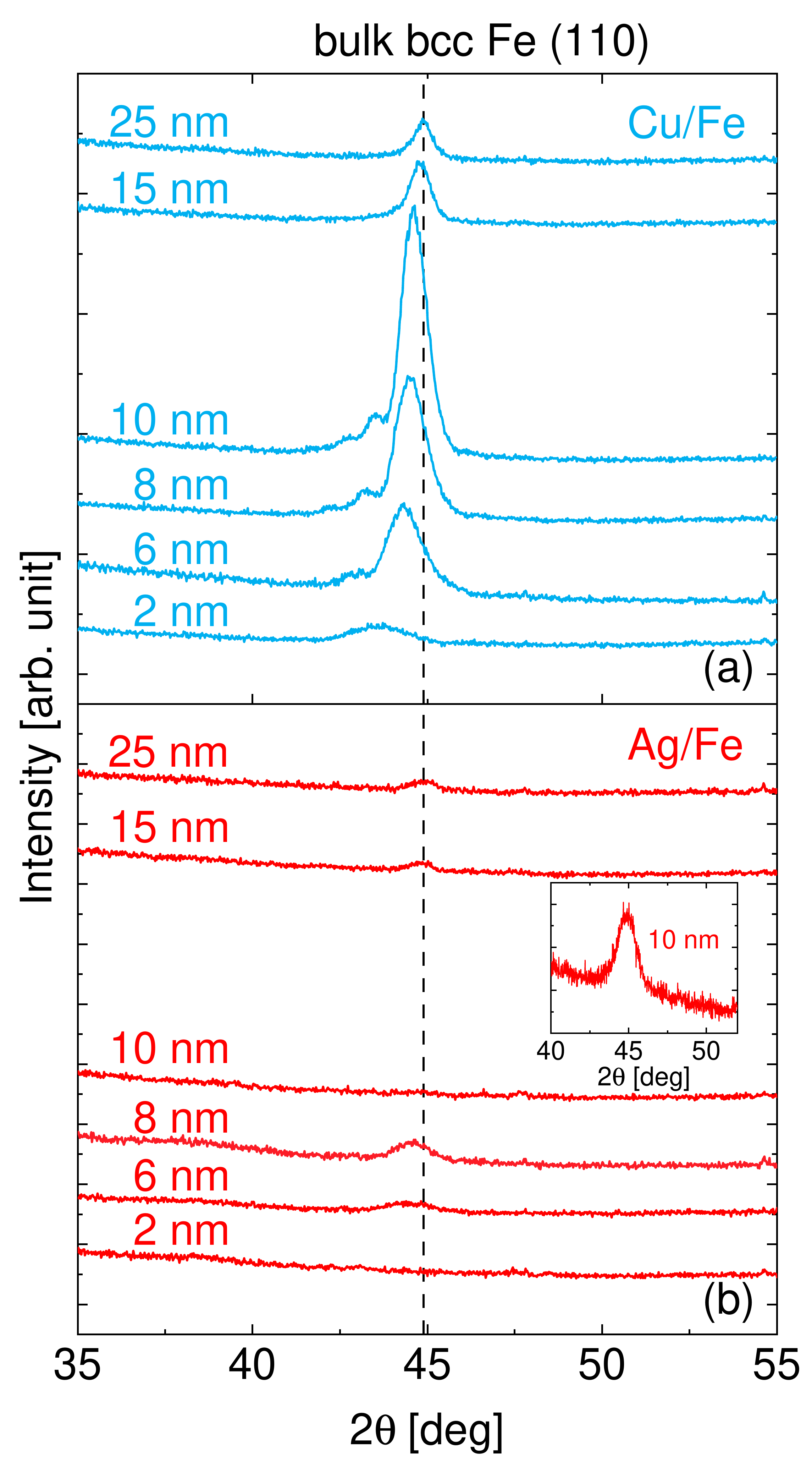}
	\includegraphics[width=0.31\linewidth]{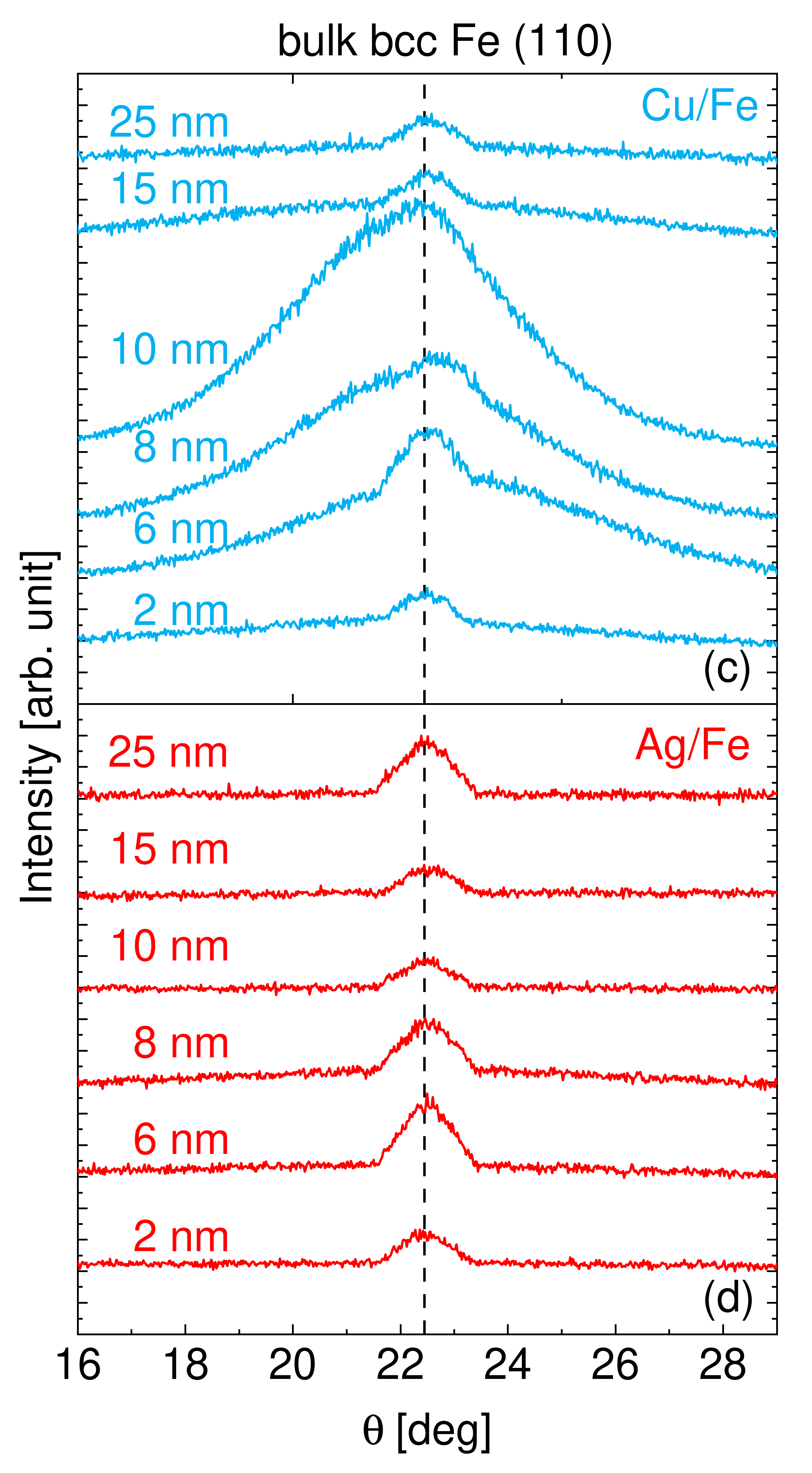}
	\includegraphics[width=0.355\linewidth]{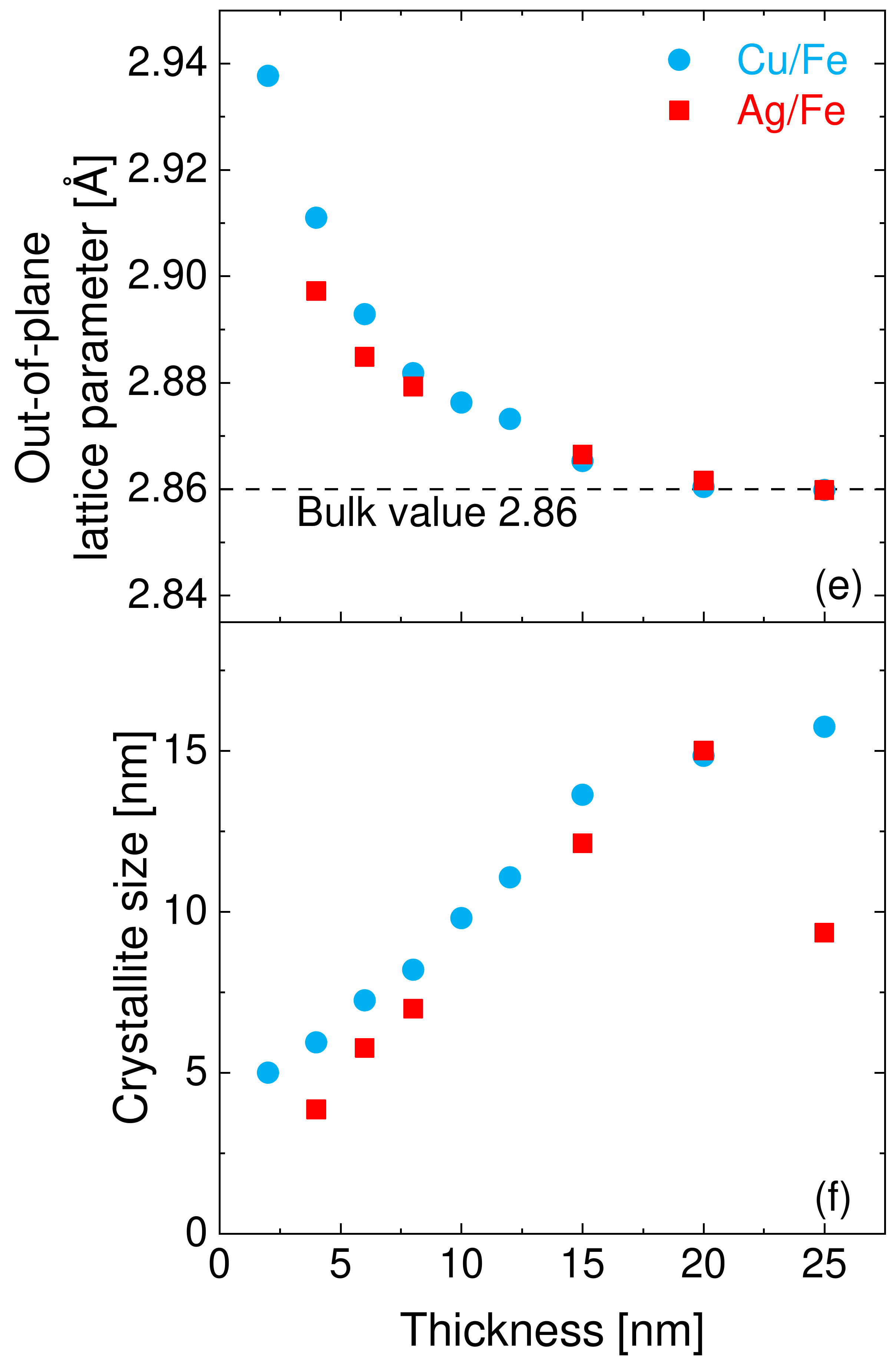}
	\caption{(Color online) $\theta$-$2\theta$ X-ray diffraction scan curves for (a) Cu/Fe (blue lines) and (b) Ag/Fe (red lines) sample series. The inset in (b) is the grazing-incidence XRD scan curve for 10 nm thick Ag/Fe film. Rocking curves for (c) Cu/Fe (blue lines) and (d) Ag/Fe (red lines) sample series. (e) Out-of-plane lattice parameter estimated via Bragg's law using the $2\theta$ value at the maximum of the tallest film diffraction peak. (f) Crystallite size estimated via the Scherrer equation using the full-width-at-half-maximum of the tallest film diffraction peak. In (e) and (f), the data for the Ag/Fe film series at a few thickness values are missing because of the absence of the bcc (110) peak in $\theta$-$2\theta$ XRD scans.}
	\label{fig:XRD}
\end{figure}

We performed x-ray diffraction (XRD) measurements to compare the structural properties of the Cu/Fe and Ag/Fe films. Figure \ref{fig:XRD}(a,b) shows symmetric $\theta$-$2\theta$ XRD scan curves for several films from both the Cu/Fe and Ag/Fe sample series. For all Cu/Fe films, the (110) body-center-cubic (bcc) peak can be observed around $2\theta=44\degree-45$\degree (Fig.~\ref{fig:XRD}(a)). This observation confirms that the Fe films grown on Cu are polycrystalline and textured, where the crystal grains predominantly possess (110)-oriented planes that are parallel to the sample surface. For Ag/Fe (Fig.~\ref{fig:XRD}(b)), the (110) bcc peak is absent or extremely weak, from which one might surmise that the Fe films grown on Ag are amorphous or only possess weak crystallographic texture. However, we find that the Ag/Fe films are, in fact, also polycrystalline with evidence of (110) texturing. In the following, we elaborate on our XRD results, first for Cu/Fe and then Ag/Fe. 

We observe evidence for a peculiar, non-monotonic trend in the microstructural properties of the Cu/Fe films. Specifically, the height of the $\theta$-$2\theta$ diffraction peak (Fig.~\ref{fig:XRD}(a)) increases with Fe film thickness up to $\approx$10 nm but then decreases at higher Fe film thicknesses. 
While we do not have a complete explanation for this peculiar nonmonotonic trend with film thickness, a closer inspection of the XRD results (Fig.~\ref{fig:XRD}) provides useful insights. 
First, the Fe film diffraction peak shifts toward a higher 2$\theta$ value with increasing film thickness. 
This signifies that thinner Fe films on Cu are strained (with the Fe crystal lattice tetragonally distorted), whereas thicker Fe films undergo structural relaxation such that the out-of-plane lattice parameter converges toward the bulk value of $\approx$2.86 \AA, as summarized in Fig.~\ref{fig:XRD}(e). 
Second, as the Fe film thickness approaches $\approx$10 nm, additional diffraction peaks appear to the left of the tall primary peak. 
We speculate that these additional peaks may originate from Fe crystals that remain relatively strained (i.e., with an out-of-plane lattice parameter larger than the bulk value), while the primary peak arises from more relaxed Fe crystals (i.e., with a lattice parameter closer to the bulk value).
The coexistence of such different Fe crystals appears to be consistent with the rocking curve measurements  (Fig.~\ref{fig:XRD}(c)), which exhibit a large broad background peak in addition to a small sharp peak for Cu/Fe films with thicknesses near $\approx$10 nm. 
As we describe in Sec.~\ref{sec:IP}, these $\approx$10 nm thick Cu/Fe samples also show distinct behaviors in extrinsic damping (highly nonlinear frequency dependence of the FMR linewidth) and static magnetization reversal (enhanced coercivity), which appear to be correlated with the peculiar microstructural properties evidenced by our XRD results.
On the other hand, it is worth noting that the estimated crystal grain size (Fig.~\ref{fig:XRD}(f)) -- derived from the width of the $\theta$-$2\theta$ diffraction peak -- does not exhibit any anomaly near the film thickness of $\approx$10 nm, but rather increases monotonically with film thickness.

Unlike the Cu/Fe films discussed above, the Ag/Fe films do not show a strong (110) bcc peak in the $\theta$-$2\theta$ XRD results.
However, the lack of pronounced peaks in the symmetric $\theta$-$2\theta$ scans does not necessarily signify that Ag/Fe is amorphous. This is because symmetric $\theta$-$2\theta$ XRD is sensitive to crystal planes that are nearly parallel to the sample surface, such that the diffraction peaks capture only the crystal planes with out-of-plane orientation with a rather small range of misalignment (within $\sim$1\degree, dictated by incident X-ray beam divergence). In fact, from \emph{asymmetric}  grazing-incidence XRD scans that are sensitive to other planes, we are able to observe a clear bcc Fe (110) diffraction peak even for Ag/Fe samples that lack an obvious diffraction peak in $\theta$-$2\theta$ scans (see e.g. inset of Fig. 1(b)). 
Furthermore, rocking curve scans (conducted with $2\theta$ fixed to the expected position of the (110) Fe film diffraction peak) provide orientation information over an angular range much wider than $\sim$1\degree. As shown in Fig.~\ref{fig:XRD}(d), a clear rocking curve peak is observed for each Ag/Fe sample, suggesting that Fe films grown on Ag are polycrystalline and (110)-textured -- albeit with the (110) crystal planes more misaligned from the sample surface compared to the Cu/Fe samples. 
The out-of-plane lattice parameters of Ag/Fe films (with discernible $\theta$-$2\theta$ diffraction film peaks) show the trend of relaxation towards the bulk value with increasing Fe thickness, similar to the Cu/Fe series. Yet, the lattice parameters for Ag/Fe at small thicknesses are systematically closer to the bulk value, possibly because Fe is less strained (i.e., better lattice matched) on Ag than on Cu. We also find that the estimation of the crystal grain size for Ag/Fe -- although made difficult by the smallness of the diffraction peak -- yields a trend comparable to Cu/Fe, as shown in Fig.~\ref{fig:XRD}(f).

\begin{figure}[!htbp]
	\centering
	\includegraphics[width=0.45\textwidth]{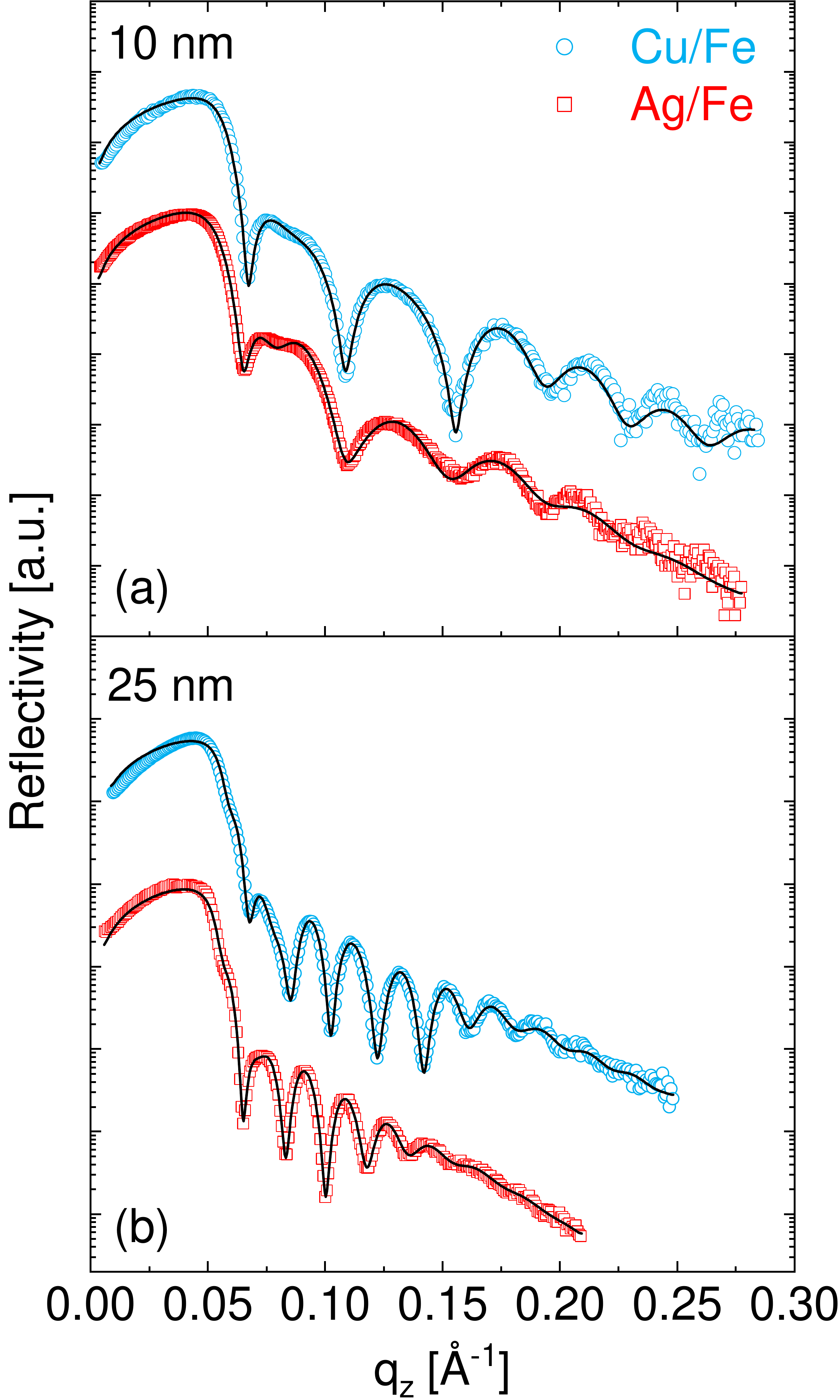}
	\caption{(Color online) X-ray reflectivity scans of 10 nm and 25 nm thick films from (a) Cu/Fe (blue circles) and (b) Ag/Fe (red squares) sample series. Black solid curves are fits to the data.}
	\label{fig:XRR}
\end{figure}

We also observe a notable difference between Cu/Fe and Ag/Fe in the properties of film interfaces, as revealed by XRR scans in Fig. \ref{fig:XRR}. The oscillation period depends inversely on the film thickness. The faster decay of the oscillatory reflectivity signal at high angles for the Ag/Fe films suggests that the Ag/Fe films may have rougher interfaces compared to the Cu/Fe films. Another interpretation of the XRR results is that the Ag/Fe interface is more diffuse than the Cu/Fe interface -- i.e., due to interfacial intermixing of Ag and Fe. By fitting the XRR results~\cite{Vignaud2019}, we estimate an average roughness (or the thickness of the diffuse interfacial layer) of $\lesssim$1 nm for the Fe layer in Cu/Fe, while it is much greater at $\approx$2-3 nm for Ag/Fe~\footnote{Here, the ``average roughness'' is the average of the roughness of the top and bottom interfaces of the Fe layer.}. 

Our structural characterization described above thus reveals key attributes of the Cu/Fe and Ag/Fe sample series. Both film series are polycrystalline, exhibit (110) texture, and have grain sizes of order film thickness. Nevertheless, there are also crucial differences between Cu/Fe and Ag/Fe. The Cu/Fe series overall exhibits stronger $\theta$-$2\theta$ diffraction peaks than the Ag/Fe series, suggesting that the (110) bcc crystal planes of Fe grown on Cu are aligned within a tighter angular range than those grown on Ag. 
Moreover, Fe grown on Cu has relatively smooth or sharp interfaces compared to Fe grown on Ag.
Although identifying the origin of such structural differences is beyond the scope of this work, Cu/Fe and Ag/Fe constitute two qualitatively distinct series of polycrystalline Fe films for exploring the influence of microstructure on magnetic relaxation.

\section{Intrinsic Gilbert Damping Probed by Out-of-plane FMR}
\label{sec:OOP}
\begin{figure}[!htbp]
	\centering
	\hspace*{0.88cm}
	\includegraphics[width=0.48\linewidth]{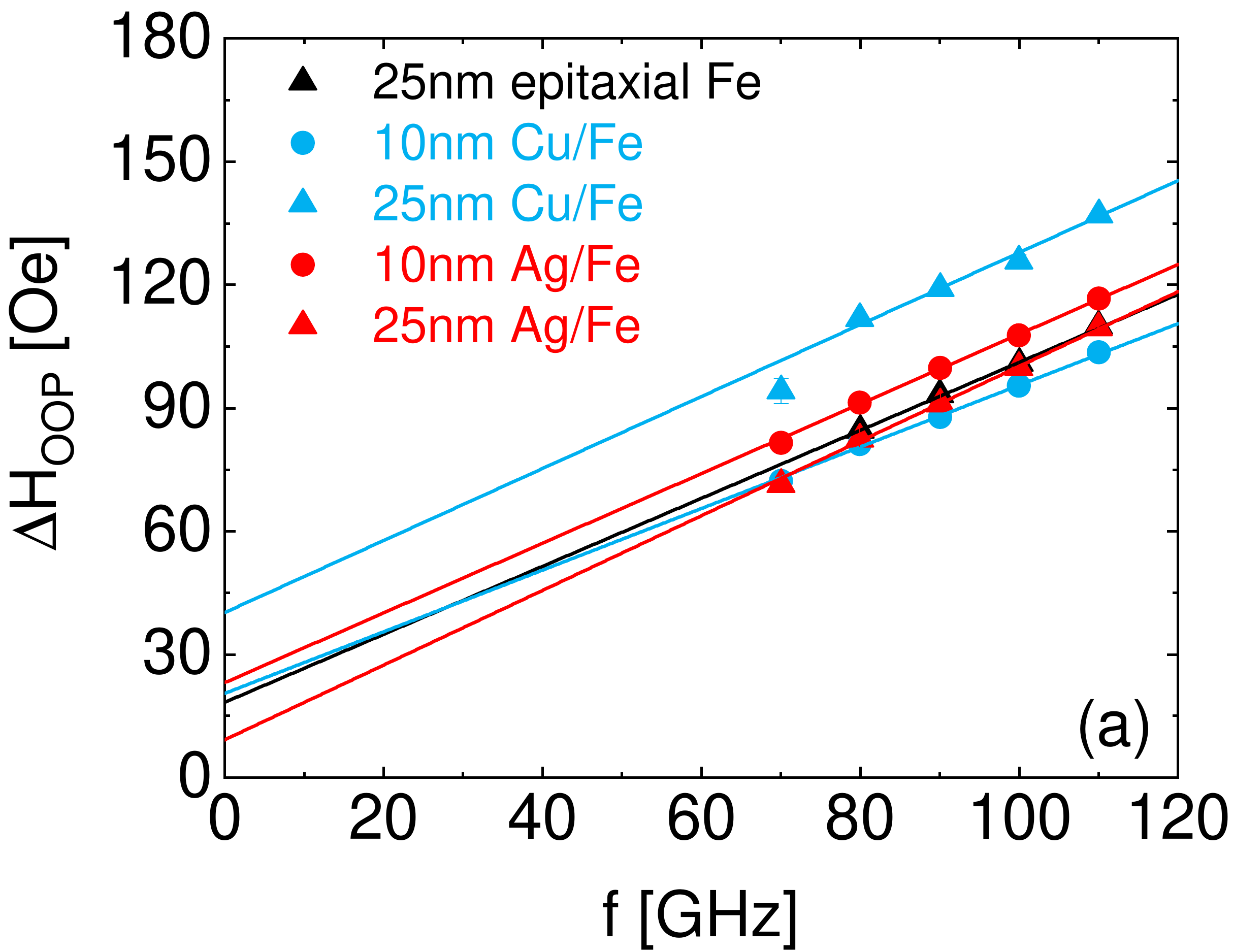}
	\includegraphics[width=0.5\linewidth]{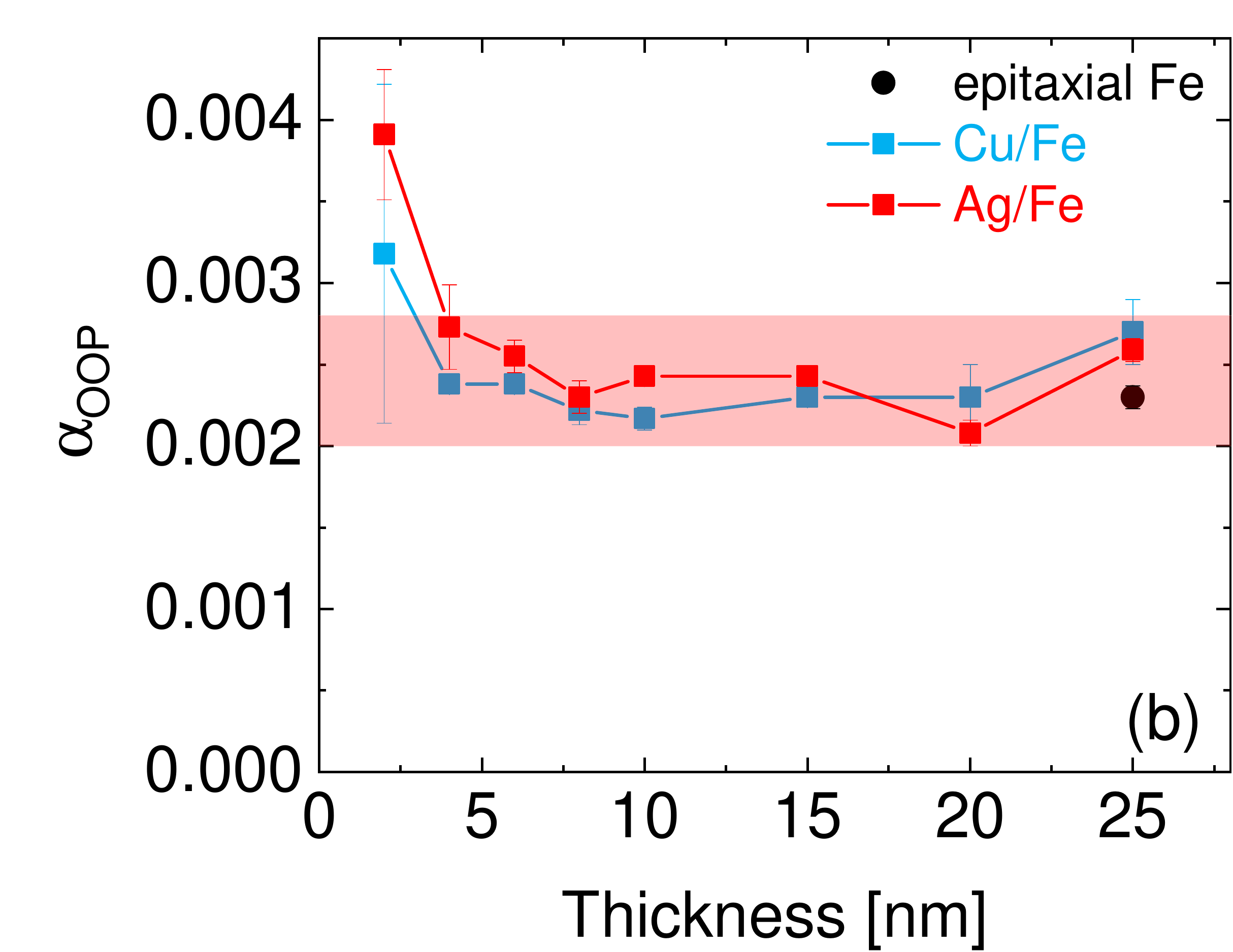}
	\caption{(Color online) (a) OOP FMR half-width-at-half-maximum linewidth $\Delta H_\mathrm{OOP}$ as a function of resonance frequency $f$. Lines correspond to fits to the data. (b) Gilbert damping parameter $\alpha_mathrm{OOP}$ extracted from OOP FMR as a function of film thickness. The red shaded area highlights the damping value range that contains data points of all films thicker than 4 nm. The data for the epitaxial Fe sample (25 nm thick Fe grown on \ce{MgAl_2O_4}) are adapted from Ref.~\cite{Khodadadi2020}. }
	\label{fig:oop}
\end{figure}

Having established the difference in structural properties between Cu/Fe and Ag/Fe, we characterize room-temperature intrinsic damping for these samples with OOP FMR measurements. The OOP geometry suppresses two-magnon scattering~\cite{McMichael2004} such that the Gilbert damping parameter can be quantified in a straightforward manner. We use a W-band shorted waveguide in a superconducting magnet, which permits FMR measurements at high fields ($\gtrsim$ 4 T) that completely magnetize the Fe films out of plane. The details of the measurement method are found in Refs.~\cite{Khodadadi2020,Smith2020}.  Figure \ref{fig:oop}(a) shows the frequency dependence of half-width-at-half-maximum (HWHM) linewidth $\Delta H_\mathrm{OOP}$ for selected thicknesses from both sample series. The linewidth data of 25 nm thick epitaxial Fe film from a previous study\cite{Khodadadi2020} is plotted in Fig. \ref{fig:oop} (a) as well. The intrinsic damping parameter can be extracted from the linewidth plot using

\begin{equation}
	\Delta H_\mathrm{OOP} = \Delta H_0 + \frac{2\pi}{\gamma}\alpha_{\mathrm{OOP}} f,
	\label{equ:GilbertDamping}
\end{equation}

\noindent where $\Delta H_0$ is the inhomogeneous broadening~\footnote{The magnitude of the inhomogenous broadening $\Delta H_0$ seen in OP FMR ranges from $\approx$10 to 50 Oe with no clear systematic dependence on Fe film thickness or seed layer material.}, $\gamma = \frac{g\mu_B}{\hbar}$ is the gyromagnetic ratio ($\gamma/2\pi \approx$ 2.9 MHz/Oe [Ref.~\footnote{$\gamma/2\pi \approx$ 2.9 MHz/Oe corresponds to a spectroscopic $g$-factor of $g\approx2.08$, in line with Ref.~\cite{Schoen2017}.}], obtained from the frequency dependence of resonance field~\cite{Smith2020}), and $\alpha_{\mathrm{OOP}}$ is the measured viscous damping parameter.
In general, $\alpha_{\mathrm{OOP}}$ can include not only intrinsic Gilbert damping, parameterized by $\alpha_{\mathrm{int}}$, but also eddy-current, radiative damping, and spin pumping contributions\cite{schoen2015radiative}, which all yield a linear frequency dependence of the linewidth. Damping due to eddy current is estimated to make up less than 10\% of the total measured damping parameter\cite{Smith2020} and is ignored here. Since we used a shorted waveguide in our setup, the radiative damping does not apply here. Spin pumping is also negligible for most of the samples here because the materials in the seed and capping layers (i.e., Ti, Cu, and Ag) possess weak spin-orbit coupling and are hence poor spin sinks~\cite{Edwards2019,Wang2014,Du2014}. We therefore proceed by assuming that the measured OOP damping parameter $\alpha_{\mathrm{OOP}}$ is equivalent to the \emph{intrinsic} Gilbert damping parameter.

The extracted damping parameter is plotted as a function of Fe film thickness in Fig. \ref{fig:oop}(b). The room-temperature damping parameters of all Fe films with thicknesses of 4-25 nm fall in the range of 0.0024 $\pm$ 0.0004, which is shaded in red in Fig. \ref{fig:oop}(b). This damping parameter range is quantitatively in line with the value reported for epitaxial Fe (black symbol in Fig.~\ref{fig:oop}(b))~\cite{Khodadadi2020}. For 2 nm thick samples, the damping parameter is larger likely due to an additional interfacial contribution~\cite{Heinrich1987,Celinski1991, ChenMankovsky2018} -- e.g., spin relaxation through interfacial Rashba spin-orbit coupling~\cite{ChenZhang2015} that becomes evident only for ultrathin Fe. The results in Fig.~\ref{fig:oop}(b) therefore indicate that the structural properties of the $\gtrsim$4 nm thick polycrystalline bcc Fe films have little influence on their intrinsic damping. 

It is remarkable that these polycrystalline Cu/Fe and Ag/Fe films -- with different thicknesses and microstructural properties (as revealed in Sec.~\ref{sec:structure}) -- exhibit essentially the same \emph{room-temperature} intrinsic Gilbert damping parameter as single-crystalline bcc Fe. This finding is qualitatively distinct from a prior report~\cite{Khodadadi2020} on intrinsic Gilbert damping in single-crystalline Fe films at \emph{cryogenic} temperature, which is sensitive to microstructural disorder. In the following, we discuss the possible differences in the mechanisms of intrinsic damping between these temperature regimes.

Intrinsic Gilbert damping in ferromagnetic metals is predominantly governed by transitions of spin-polarized electrons between electronic states, within a given electronic band (intraband scattering) or in different electronic bands (interband scattering) near the Fermi level \cite{Gilmore2007}. 
For Fe, previous studies~\cite{Khodadadi2020,Gilmore2007,Gilmore2007a} indicate that intraband scattering tends to dominate at low temperature where the electronic scattering rate is low (e.g., $\sim$10$^{13}$ s$^{-1}$); by contrast, interband scattering likely dominates at room temperature where the electronic scattering rate is higher (e.g., $\sim$10$^{14}$ s$^{-1}$). 
According to our results (Fig.~\ref{fig:oop}(b)), intrinsic damping at room temperature is evidently unaffected by the variation in the structural properties of the Fe films. Hence, the observed intrinsic damping is mostly governed by the electronic band structure \emph{within the Fe grains}, such that disorder in grain boundaries or film interfaces has minimal impact. 

The question remains as to why interband scattering at room temperature leads to Gilbert damping that is insensitive to microstructural disorder, in contrast to intraband scattering at low temperature yielding damping that is quite sensitive to microstructure~\cite{Khodadadi2020}. This distinction may be governed by what predominantly drives electronic scattering -- specifically, defects (e.g., grain boundaries, rough or diffuse interfaces) at low temperature, as opposed to phonons at high temperature. That is, the dominance of phonon-driven scattering at room temperature may effectively diminish the roles of microstructural defects in Gilbert damping. Future experimental studies of temperature-dependent damping in polycrystalline Fe films may provide deeper insights.
Regardless of the underlying mechanisms, the robust consistency of $\alpha_\mathrm{OOP}$ (Fig.~\ref{fig:oop}(b)) could be an indication that the intrinsic Gilbert damping parameter at a sufficiently high temperature is a local property of the ferromagnetic metal, possibly averaged over the ferromagnetic exchange length of just a few nm \cite{Abo2013} that is comparable or smaller than the grain size. 
In this scenario, the impact on damping from grain boundaries would be limited in comparison to the contributions to damping within the grains. 

Moreover, the misalignment of Fe grains evidently does not have much influence on the intrinsic damping. This is reasonable considering that intrinsic Gilbert damping is predicted to be nearly isotropic in Fe at sufficiently high electronic scattering rates~\cite{Gilmore2010} -- e.g., $\sim$10$^{14}$ s$^{-1}$ at room temperature where interband scattering is expected to be dominant~\cite{Khodadadi2020,Gilmore2007,Gilmore2007a}.
It is also worth emphasizing that $\alpha_\mathrm{OOP}$ remains unchanged for Fe films of various thicknesses with different magnitudes of strain (tetragonal distortion, as evidenced by the variation in the out-of-plane lattice parameter in Fig.~\ref{fig:XRD}(e)).
Strain in Fe grains is not expected to impact the intrinsic damping, as Ref.~\cite{Khodadadi2020} suggests that strain in bcc Fe does not significantly alter the band structure near the Fermi level. Thus, polycrystalline Fe films exhibit essentially the same magnitude of room-temperature intrinsic Gilbert damping as epitaxial Fe, as long as the grains retain the bcc crystal structure.     

The observed invariance of intrinsic damping here is quite different from the recent study of polycrystalline \ce{Co_{25}Fe_{75}} alloy films\cite{Edwards2019}, reporting a decrease in intrinsic damping with increasing structural disorder. This inverse correlation between intrinsic damping and disorder in Ref.~\cite{Edwards2019} is attributed to the dominance of intraband scattering, which is inversely proportional to the electronic scattering rate. It remains an open challenge to understand why the room-temperature intrinsic Gilbert damping of some ferromagnetic metals might be more sensitive to structural disorder than others.


\section{Extrinsic Magnetic Relaxation Probed by In-plane FMR}
\label{sec:IP}

Although we have shown via OOP FMR in Sec.~\ref{sec:OOP} that intrinsic Gilbert damping is essentially independent of the structural properties of the Fe films, it might be expected that microstructure has a pronounced impact on \emph{extrinsic} magnetic relaxation driven by two-magnon scattering, which is generally present in IP FMR. IP magnetized films are more common in device applications than OOP magnetized films, since the shape anisotropy of thin films tends to keep the magnetization in the film plane. What governs the performance of such magnetic devices (e.g., quality factor\cite{Hou2019,Li2019}) may not be the intrinsic Gilbert damping parameter but the total FMR linewidth. Thus, for many magnetic device applications, it is essential to understand the contributions to the IP FMR linewidth. 

\begin{figure}[!htbp]
	\centering
	\includegraphics[width=0.5\linewidth]{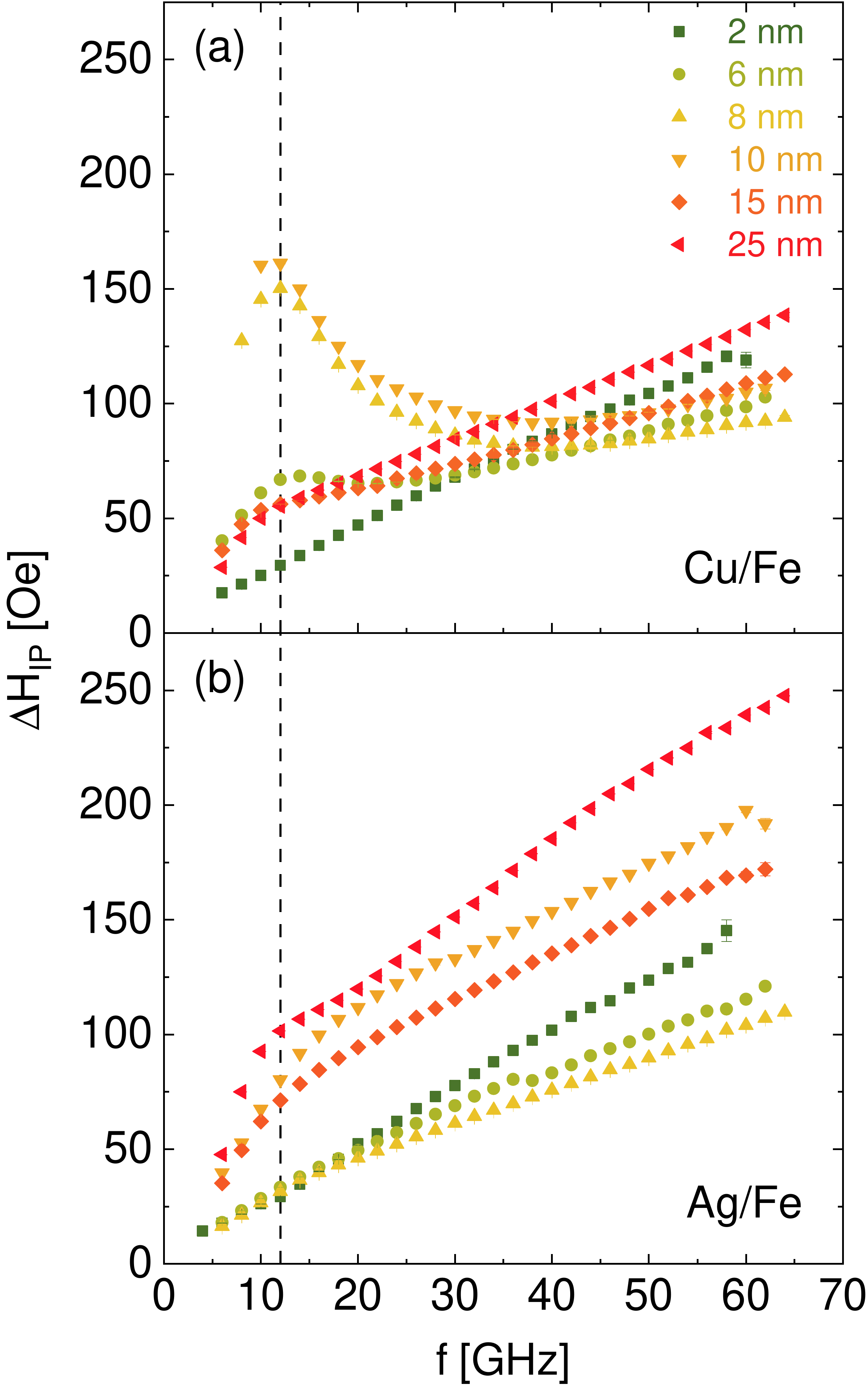}
	\caption{(Color online) IP FMR half-width-at-half-maximum linewidth $\Delta H_{\mathrm{IP}}$ as a function of resonance frequency $f$ for (a) Cu/Fe and (b) Ag/Fe. The vertical dashed line at 12 GHz highlights the hump in linewidth vs frequency seen for many of the samples.}
	\label{fig:ip}
\end{figure}

IP FMR measurements have been performed using a coplanar-waveguide-based spectrometer, as detailed in Refs.~\cite{Khodadadi2020, Smith2020}. Examples of the frequency dependence of IP FMR linewidth are shown in Fig. \ref{fig:ip}. In contrast to the linear frequency dependence that arises from intrinsic Gilbert damping in Fig. \ref{fig:oop}(a), a nonlinear hump is observed for most of the films in the vicinity of $\approx$12 GHz. In some films, e.g., 10 nm thick Cu/Fe film, the hump is so large that its peak even exceeds the linewidth at the highest measured frequency. Similar nonlinear IP FMR linewidth behavior has been observed in Fe alloy films\cite{Kalarickal2008} and epitaxial Heusler films\cite{Peria2020} in previous studies, where two-magnon scattering has been identified as a significant contributor to the FMR linewidth. Therefore, in the following, we attribute the nonlinear behavior to two-magnon scattering. 

\begin{figure}[!htbp]
	\centering
	\includegraphics[width=0.5\linewidth]{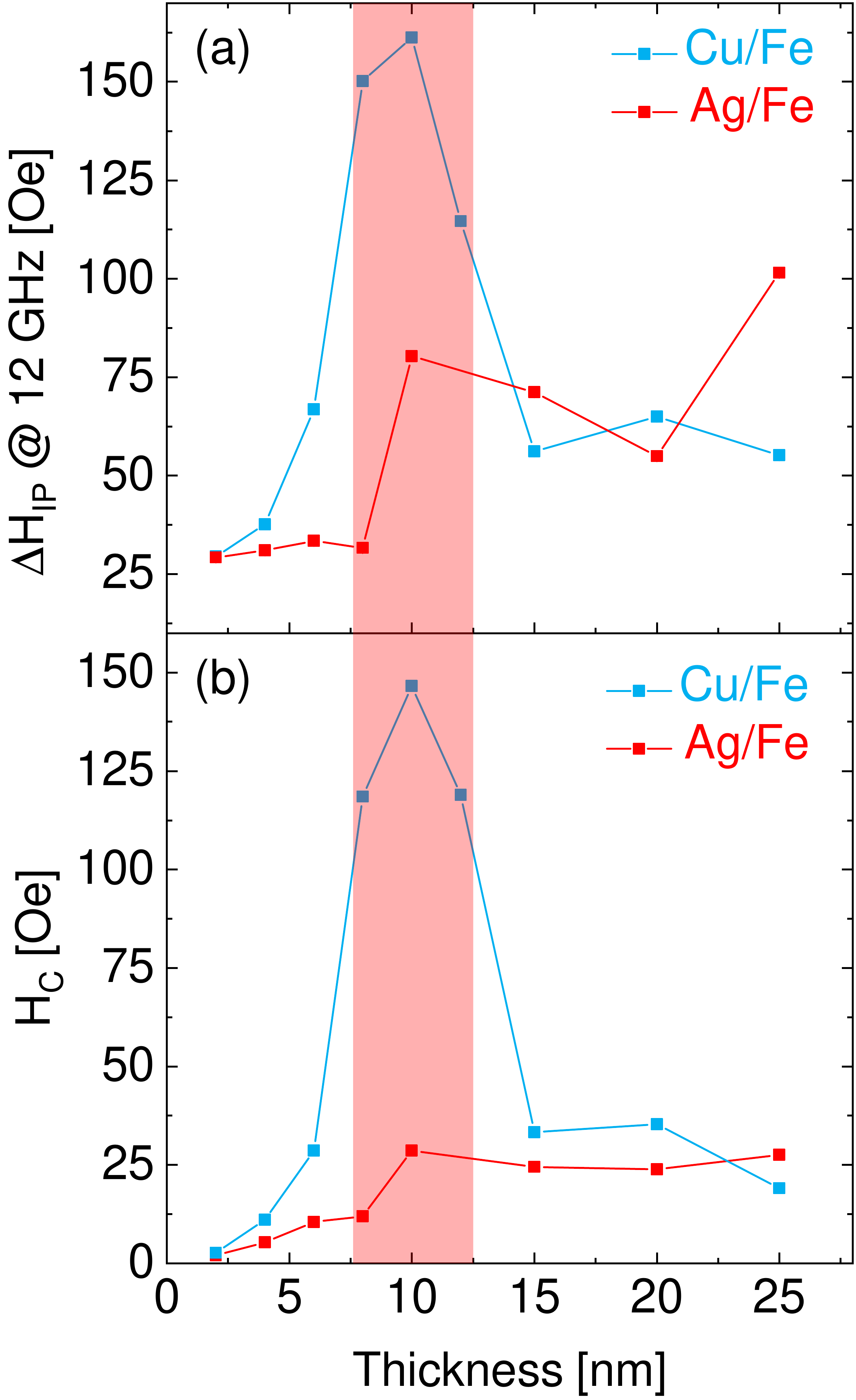}
	\caption{(Color online) (a) IP FMR half-width-at-half-maximum linewidth at 12 GHz -- approximately where the maximum (``hump'') in linewidth vs frequency is seen (see Fig.~\ref{fig:ip}) -- as a function of film thickness for both Cu/Fe and Ag/Fe. (b) Coercivity $H_c$ as a function of film thickness for both Cu/Fe and Ag/Fe. The red shaded area highlights thickness region where the Cu/Fe sample series show a peak behavior in both plots.}
	\label{fig:Hc_dH12}
\end{figure}

To gain insight into the origin of two-magnon scattering, we plot the linewidth at 12 GHz -- approximately where the hump is seen in Fig.~\ref{fig:ip} -- against the Fe film thickness in Fig. \ref{fig:Hc_dH12}(a). We do not observe a monotonic decay in the linewidth with increasing thickness that would result from two-magnon scattering of interfacial origin \cite{Azevedo2000}. Rather, we observe a non-monotonic thickness dependence in Fig. \ref{fig:Hc_dH12}(a), which indicates that the observed two-magnon scattering originates within the bulk of the films. We note that Ag/Fe with greater interfacial disorder (see Sec.~\ref{sec:structure}) exhibits weaker two-magnon scattering than Cu/Fe, particularly in the lower thickness regime ($\lesssim$10 nm). This observation further corroborates that the two-magnon scattering here is not governed by the interfacial roughness of Fe films. The contrast between Cu/Fe and Ag/Fe also might appear counterintuitive, since two-magnon scattering is induced by defects and hence might be expected to be stronger for more ``defective" films (i.e., Ag/Fe in this case). The counterintuitive nature of the two-magnon scattering here points to more subtle mechanisms at work.  

To search for a possible correlation between static magnetic properties and two-magnon scattering, we have performed vibrating sample magnetometry (VSM) measurements with a Microsense EZ9 VSM. Coercivity extracted from VSM measurements is plotted as a function of film thickness in Fig. \ref{fig:Hc_dH12}(b), which shows a remarkably close correspondence with linewidth vs thickness (Fig. \ref{fig:Hc_dH12}(a)). In particular, a pronounced peak in coercivity is observed for Cu/Fe around 10 nm, corresponding to the same thickness regime where 
the 12 GHz FMR linewidth for Cu/Fe is maximized. Moreover, the 10 nm Cu/Fe sample (see Sec.~\ref{sec:structure}) exhibits a tall, narrow bcc (110) diffraction peak, which suggests that its peculiar microstructure plays a possible role in the large two-magnon scattering and coercivity (e.g., via stronger domain wall pinning).    

While the trends shown in Fig.~\ref{fig:Hc_dH12} provide some qualitative insights, we now attempt to quantitatively analyze the frequency dependence of FMR linewidth for the Cu/Fe and Ag/Fe films. We assume that the Gilbert damping parameter for IP FMR is equal to that for OOP FMR, i.e., $\alpha_\mathrm{IP} = \alpha_\mathrm{OOP}$. This assumption is physically reasonable, considering that Gilbert damping is theoretically expected to be isotropic in Fe films near room temperature~\cite{Gilmore2010}. While a recent study has reported anisotropic Gilbert damping that scales quadratically with magnetostriction~\cite{Peria2021}, this effect is likely negligible in elemental Fe whose magnetostriction is several times smaller~\cite{Clark2003, Summers2007} than that of the Fe$_{0.7}$Ga$_{0.3}$ alloy in Ref.~\cite{Peria2021}. 

\begin{figure}[!htbp]
	\centering
	\includegraphics[width=0.75\linewidth]{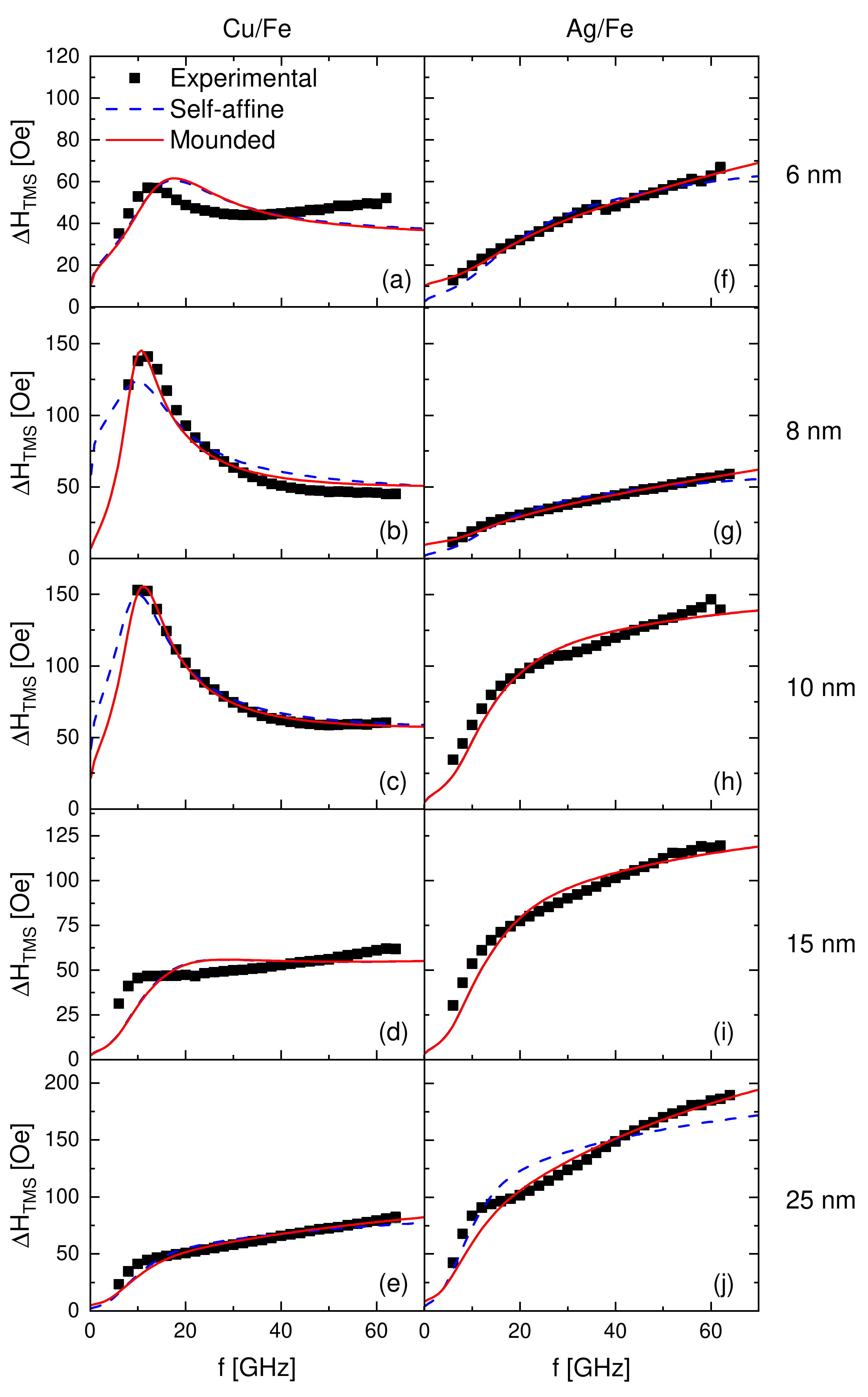}
	\caption{(Color online) Extrinsic two-magnon scattering linewidth $\Delta H_{\mathrm{TMS}}$ vs frequency $f$ and fitted curves for 6, 8, 10, 15, and 25 nm Cu/Fe and Ag/Fe films. Black squares represent experimental FMR linewidth data. Dashed blue and solid red curves represent the fitted curves using correlation functions proposed for modeling self-affine and mounded surfaces, respectively. In (d), (e), (h), (i), dashed blue curves overlap with solid red curves.}
	\label{fig:ip_fit}
\end{figure}

Thus, from the measured IP linewidth $\Delta H_{\mathrm{IP}}$, the extrinsic two-magnon scattering linewidth $\Delta H_{\mathrm{TMS}}$ can be obtained by
\begin{equation}
    \Delta H_{\mathrm{TMS}} = \Delta H_{\mathrm{IP}} - \frac{2\pi}{\gamma}\alpha_\mathrm{IP},
\end{equation}
where $\frac{2\pi}{\gamma}\alpha_\mathrm{IP} $ is the Gilbert damping contribution. Figure \ref{fig:ip_fit} shows the obtained $\Delta H_{\mathrm{TMS}}$ and fit attempts using the ``grain-to-grain'' two-magnon scattering model developed by McMicheal and Krivosik\cite{McMichael2004}. This model captures the inhomogeneity of the effective internal magnetic field in a film consisting of many magnetic grains. The magnetic inhomogeneity can arise from the distribution of magnetocrystalline anisotropy field directions associated with the randomly oriented crystal grains~\cite{Kalarickal2008}. In this model the two-magnon scattering linewidth $\Delta H_{\mathrm{TMS}}$ is a function of the Gilbert damping parameter $\alpha_\mathrm{IP}$, the effective anisotropy field $H_a$ of the randomly oriented grain, and the correlation length $\xi$ within which the effective internal magnetic field is correlated. Further details for computing $\Delta H_{\mathrm{TMS}}$ are provided in the Appendix and Refs.~\cite{McMichael2004,Kalarickal2008,Peria2020}. As we have specified above, $\alpha_\mathrm{IP}$ is set to the value derived from OOP FMR results (i.e., $\alpha_\mathrm{OOP}$ in Fig.~\ref{fig:oop}(b)). This leaves $\xi$ and $H_a$ as the only free parameters in the fitting process. 

The modeling results are dependent on the choice of the correlation function $C(\mathbf{R})$, which captures how the effective internal magnetic field is correlated as a function of lateral distance $\mathbf{R}$ in the film plane. We first show results obtained with a simple exponentially decaying correlation function, as done in prior studies of two-magnon scattering~\cite{McMichael2004,Kalarickal2008,Peria2020}, i.e., 
\begin{equation}
\label{eq:selfaffine}
    C(\mathbf{R})=\exp\left(-\frac{|\mathbf{R}|}{\xi}\right).
\end{equation}
Equation~\ref{eq:selfaffine} has the same form as the simplest correlation function used to model rough topographical surfaces (when they are assumed to be ``self-affine'')~\cite{pelliccione2008evolution}. Fit results with Eq.~(\ref{eq:selfaffine}) are shown in dashed blue curves in Fig. \ref{fig:ip_fit}. For most samples, the fitted curve does not reproduce the experimental data quantitatively. Moreover, the fitted values of $\xi$ and $H_a$ often reach physically unrealistic values, e.g., with $H_a > 10^4$ Oe and $\xi < 1$ nm  (see Table~\ref{tab:fit}). These results suggest that the model does not properly capture the underlying physics of two-magnon scattering in our samples. 


A possible cause for the failure to fit the data is that the simple correlation function (Eq.~\ref{eq:selfaffine}) is inadequate. We therefore consider an alternative correlation function by again invoking an analogy between the spatially varying height of a rough surface~\cite{pelliccione2008evolution} and the spatially varying effective internal magnetic field in a film. Specifically, we apply a correlation function (i.e., a special case of Eq.~(4.3)  in Ref.~\cite{pelliccione2008evolution} where short-range roughness $\alpha=1$) for the so-called ``mounded surface,'' which incorporates the average distance $\lambda$ between peaks in topographical height (or, analogously, effective internal magnetic field):  
\begin{equation}
\label{eq:mounded}
    C(\mathbf{R})=\frac{\sqrt{2}|\mathbf{R}|}{\xi}K_{1}\left(\frac{\sqrt{2}|\mathbf{R}|}{\xi}\right)J_0\left(\frac{2\pi|\mathbf{R}|}{\lambda}\right),
\end{equation}
where $J_0$ and $K_1$ are the Bessel function of the first kind of order zero and the modified Bessel function of the second kind of order one, respectively. This oscillatory decaying function is chosen because its Fourier transform (see Appendix) does not contain any transcendental functions, which simplifies the numerical calculation. 
We also stress that while Eq.~(\ref{eq:mounded}) in the original context (Ref.~\cite{pelliccione2008evolution}) was used to model topographical roughness, we are applying Eq.~(\ref{eq:mounded}) in an attempt to model the spatial fluctuations (``roughness") of the effective internal magnetic field -- rather than the roughness of the film topography.

The fitted curves using the model with Eq.~(\ref{eq:mounded}) are shown in solid red curves in Fig. \ref{fig:ip_fit}. Fit results for some samples show visible improvement, although this is perhaps not surprising with the introduction of $\lambda$ as an additional free parameter. Nevertheless, the fitted values of $H_a$ or $\lambda$ still diverge to unrealistic values of $>10^4$ Oe or $>10^4$ nm in some cases (see Table~\ref{tab:fit}), which means that the new correlation function (Eq.~(\ref{eq:mounded})) does not fully reflect the meaningful underlying physics of our samples either. More detailed characterization of the microstructure and inhomogeneities, e.g., via synchrotron x-ray and neutron scattering, could help determine the appropriate correlation function. It is also worth pointing out that for some samples (e.g. 15 nm Cu/Fe and Ag/Fe films), essentially identical fit curves are obtained regardless of the correlation function. This is because when $\lambda \gg \xi$, the Fourier transform of Eq.~(\ref{eq:mounded}) has a very similar form as the Fourier transform of Eq.~(\ref{eq:selfaffine}), as shown in the Appendix. In such cases, the choice of the correlation function has almost no influence on the behavior of the two-magnon scattering model in the fitting process. 
\begin{table}[htbp]
	\caption{Summary of IP FMR linewidth fit results. Note the divergence to physically unreasonable values in many of the results. Standard error is calculated using equation $\sqrt{\mathrm{SSR}/\mathrm{DOF}\times \mathrm{diag}(\mathbf{COV})}$, where SSR stands for the sum of squared residuals, DOF stands for degrees of freedom, and $\mathbf{COV}$ stands for the covariance matrix.}
	\centering
	\renewcommand\cellset{\renewcommand{\arraystretch}{0.7}}
	\begin{tabular}{c@{\hskip 0.5cm}c@{\hskip 0.25cm}|@{\hskip 0.25cm}c@{\hskip 0.5cm}c@{\hskip 0.25cm}|@{\hskip 0.25cm}c@{\hskip 0.5cm}c@{\hskip 0.5cm}c}
		\toprule[1pt]
		&&\multicolumn{2}{c@{\hskip 0.25cm}|@{\hskip 0.25cm}}{Self-affine}&\multicolumn{3}{c}{Mounded}\\
		\makecell{Sample\\Series}&\makecell{Thickness\\(nm)}&\makecell{$\xi$ \\ (nm)}&\makecell{$H_a$\\ (Oe)}&\makecell{$\xi$ \\ (nm)}&\makecell{$H_a$\\ (Oe)}&\makecell{$\lambda$ \\ (nm)}\\
		\cmidrule[0.5pt](lr){1-7}
		\multirow{5}{*}{Cu/Fe}&6&70 $\pm$ 10&170 $\pm$ 10&80 $\pm$ 90&24 $\pm$ 3&\textgreater\num{1e4} \\
		&8&200 $\pm$ 100&150 $\pm$ 20&700 $\pm$ 1000&25 $\pm$ 2&900 $\pm$ 100\\
		&10&140 $\pm$ 40&200 $\pm$ 20&160 $\pm$ 50&33 $\pm$ 1&800 $\pm$ 200\\
		&15&9 $\pm$ 2&800 $\pm$ 100&10 $\pm$ 20&100 $\pm$ 80&\textgreater\num{1e4}\\
		&25&0 $\pm$ 5&\textgreater\num{1e4}&60 $\pm$ 30&\textgreater\num{1e4}&10.41 $\pm$ 0.01\\
		\cmidrule[0.5pt](lr){1-7}
		\multirow{5}{*}{Ag/Fe}&6&0 $\pm$ 40&\textgreater\num{1e4}&150 $\pm$ 40&\textgreater\num{1e4}&11.7 $\pm$ 0.7\\
		&8&0 $\pm$ 30&\textgreater\num{1e4}&170 $\pm$ 50&\textgreater\num{1e4}&12 $\pm$ 4\\
		&10&6 $\pm$ 1&1500 $\pm$ 300&8 $\pm$ 40&200 $\pm$ 500&\textgreater\num{1e4}\\
		&15&2 $\pm$ 2&4000 $\pm$ 3000&3 $\pm$ 9&500 $\pm$ 900&\textgreater\num{6e3}\\
		&25&0 $\pm$ 6&\textgreater\num{1e4}&140 $\pm$ 50&\textgreater\num{1e4}&15 $\pm$ 6\\
		\bottomrule[1pt]
	\end{tabular}
	\label{tab:fit}
\end{table}

\section{Summary}
We have examined room-temperature intrinsic and extrinsic damping in two series of polycrystalline Fe thin films with distinct structural properties. Out-of-plane FMR measurements confirm constant intrinsic Gilbert damping of $\approx$ 0.0024, essentially independent of film thickness and structural properties. This finding implies that intrinsic damping in Fe at room temperature is predominantly governed by the crystalline and electronic band structures within the grains, rather than scattering at grain boundaries or film surfaces. The results from in-plane FMR, where extrinsic damping (i.e., two-magnon scattering) plays a significant role, are far more nuanced. The conventional grain-to-grain two-magnon scattering model fails to reproduce the in-plane FMR linewidth data with physically reasonable parameters -- pointing to the need to modify the model, along with more detailed characterization of the film microstructure. Our experimental findings advance the understanding of intrinsic Gilbert damping in polycrystalline Fe, while motivating further studies to uncover the mechanisms of extrinsic damping in structurally disordered thin films. 

\section*{Acknowledgments}
S.W. acknowledges support by the ICTAS Junior Faculty Program. D.A.S. and S.E. acknowledge support by the National Science Foundation, Grant No. DMR-2003914. P. N. would like to acknowledge support through NASA Grant NASA CAN80NSSC18M0023. A. R. would like to acknowledge support through the Defense Advanced Research Project Agency (DARPA) program on Topological Excitations in Electronics (TEE) under Grant No. D18AP00011. This work was supported by NanoEarth, a member of National Nanotechnology Coordinated Infrastructure (NNCI), supported by NSF (ECCS 1542100).

\appendix
\section{Details of the Two-Magnon Scattering Model}
In the model developed by McMichael and Krivosik, the two-magnon scattering contribution $\Delta H_\mathrm{TMS}$ to the FMR linewidth is given by\cite{McMichael2004,Kalarickal2008,Peria2020}

\begin{equation}
    \Delta H_{\mathrm{TMS}} = \frac{\gamma^2H_a^2}{2\pi P_A(\omega)}\int \Lambda_{0k}C_k(\xi)\delta_\alpha(\omega-\omega_k)\dd^2k
    \label{equ:tms}
\end{equation}

\noindent where $\xi$ is correlation length, $H_a$ is the effective anisotropy field of the randomly oriented grain. ${P_A(\omega)} = \frac{\partial \omega}{\partial H}\big|_{H = H_{\mathrm{FMR}}} = \sqrt{1+(\frac{4\pi M_s}{2\omega/\gamma})^2}$ accounts for the conversion between the frequency and field swept linewidth. $\Lambda_{0k}$ represents the averaging of the anisotropy axis fluctuations over the sample. It also takes into account the ellipticity of the precession for both the uniform FMR mode and the spin wave mode\cite{Kalarickal2008}. The detailed expression of $\Lambda_{0k}$ can be found in the Appendix of Ref. \cite{Kalarickal2008}. The coefficients in the expression of $\Lambda_{0k}$ depend on the type of anisotropy of the system. Here, we used first-order cubic anisotropy for bcc Fe. $\delta_\alpha(\omega-\omega_k)$ selects all the degenerate modes, where $\omega$ represents the FMR mode frequency and $\omega_k$ represents the spin wave mode frequency. The detailed expression of $\omega_k$ can be found in Ref. \cite{McMichael2004}. In the ideal case where Gilbert damping is 0, $\delta_\alpha$ is the Dirac delta function. For a finite damping, $\delta_\alpha(\omega_0-\omega_k)$ is replaced by a Lorentzian function $\frac{1}{\pi}\frac{(\alpha_\mathrm{IP} \omega_k/\gamma) \partial\omega/\partial H}{(\omega_k-\omega)^2+[(\alpha_\mathrm{IP} \omega_k/\gamma) \partial\omega/\partial H]^2}$, which is centered at $\omega$ and has the width of $(2\alpha_\mathrm{IP} \omega_k/\gamma) \partial\omega/\partial H$. 

\begin{figure}
    \centering
    \includegraphics[width=0.7\linewidth]{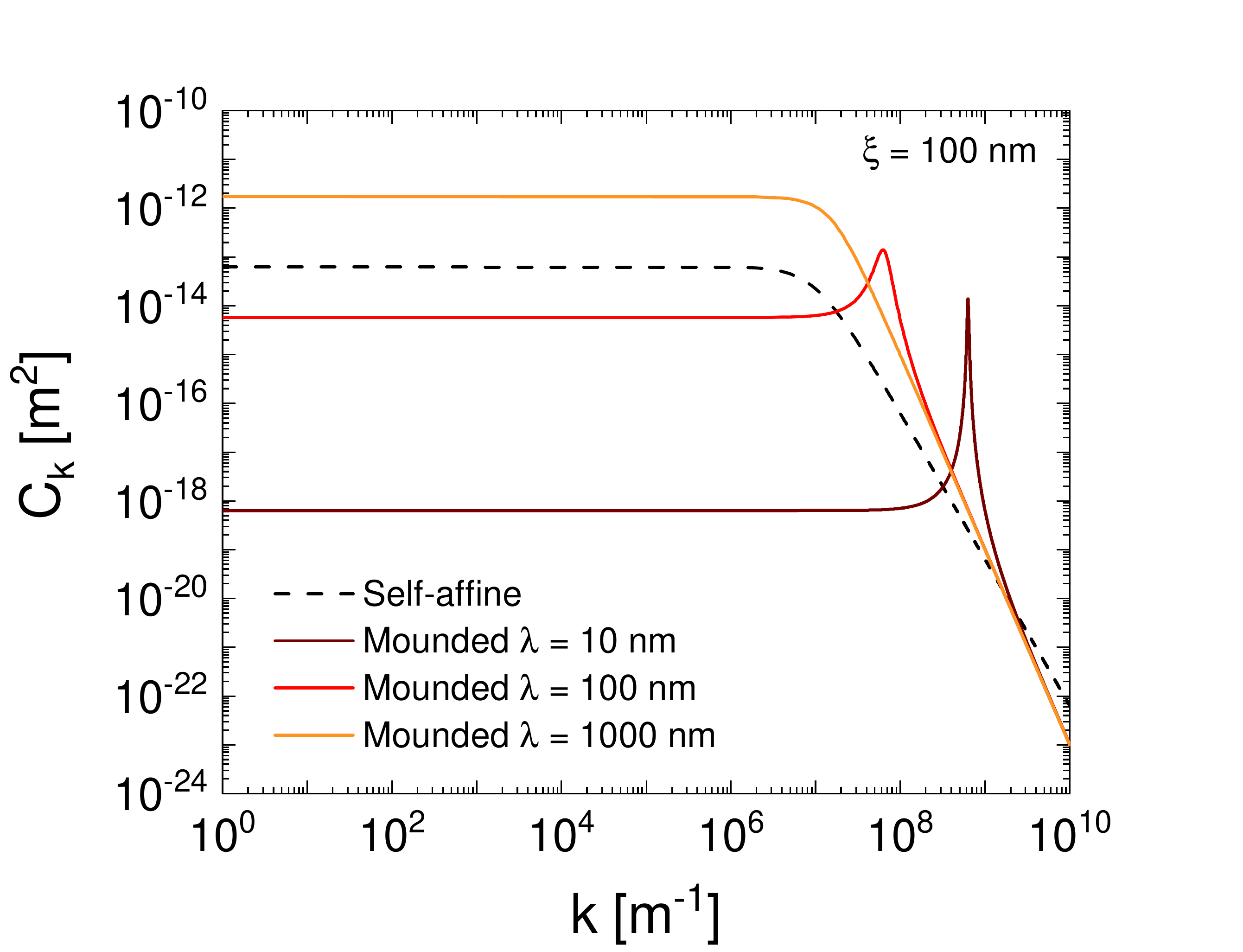}
    \caption{Fourier transform of correlation function for mounded surfaces as a function of wavenumber $k$ for three different $\lambda$ values. Fourier transform of correlation function for self-affine surfaces as a function of $k$ is also included for comparison purpose. $\xi$ is set as 100 nm for all curves.}
    \label{fig:Ck}
\end{figure}

Finally, $C_k(\xi)$ (or $C_k(\xi,\lambda)$) is the Fourier transform of the grain-to-grain internal field correlation function, Eq.~(\ref{eq:selfaffine}) (or Eq.~(\ref{eq:mounded})). For the description of magnetic inhomogeneity analogous to the simple self-affine topographical surface~\cite{pelliccione2008evolution}, the Fourier transform of the correlation function, Eq.~(\ref{eq:selfaffine}), is
\begin{equation}
    C_k(\xi)= \frac{2\pi \xi^2}{[1+(k\xi)^2]^\frac{3}{2}},
    \label{eq:selfaffine_FT}
\end{equation}
as also used in Refs. \cite{McMichael2004,Kalarickal2008,Peria2020}. For the description analogous to the mounded surface, the Fourier transform of the correlation function, Eq.~(\ref{eq:mounded}), is\cite{pelliccione2008evolution}
\begin{equation}
    C_k(\xi,\lambda) = \frac{8\pi^3\xi^2\left(1+\frac{2\pi^2\xi^2}{\lambda^2}+\frac{\xi^2}{2}k^2\right)}{\left[\left(1+\frac{2\pi^2\xi^2}{\lambda^2}+\frac{\xi^2}{2}k^2\right)^2-\left(\frac{2\pi\xi^2}{\lambda}k\right)^2\right]^{3/2}}.
    \label{eq:mounded_FT}
\end{equation}
When $\lambda \gg \xi$, Eq.~(\ref{eq:mounded_FT}) becomes
\begin{equation}
    C_k(\xi) \approx \frac{8\pi^3\xi^2}{\left(1+\frac{\xi^2}{2}k^2\right)^2},
\end{equation}
which has a similar form as Eq.~(\ref{eq:selfaffine_FT}). This similarity can also be demonstrated graphically. Figure \ref{fig:Ck} plots a self-affine $C_k$ curve (Eq.~(\ref{eq:selfaffine_FT})) at $\xi = 100$ nm and three mounded $C_k$ curves (Eq.~(\ref{eq:mounded_FT})) at $\lambda =$ 10, 100, 1000 nm. $\xi$ in mounded $C_k$ curves is set as 100 nm as well. It is clearly shown in Fig. \ref{fig:Ck} that when $\lambda = 1000$ nm, the peak appearing in $\lambda = 10$ and 100 nm mounded $C_k$ curves disappears and the curve shape of mounded $C_k$ resembles that of self-affine $C_k$.

The hump feature in Fig. \ref{fig:ip} is governed by both $\delta_\alpha$ and $C_k$ (see Eq.~\ref{equ:tms}). $\delta_\alpha$ has the shape of $\infty$ in reciprocal space ($k$ space), as shown in our videos in the Supplemental Material as well as Fig. 5(b) of Ref. \cite{Peria2020} and Fig 2 (b) of Ref. \cite{McMichael2004}. The size of the contour of the degenerated spin wave modes in $k$ space increases as the microwave frequency $f$ increases, which means the number of available degenerate spin wave modes increases as $f$ increases. As shown in Fig. \ref{fig:Ck}, self-affine $C_k$ is nearly constant with the wavenumber $k$ until $k$ reaches $\sim$$1/\xi$. This suggests that the system becomes effectively more uniform (i.e. weaker inhomogeneous perturbation) when the length scale falls below the characteristic correlation length $\xi$ (i.e., $k > 1/\xi$). Because inhomogeneities serve as the scattering centers of two-magnon scattering process, degenerate spin wave modes with $k > 1/\xi$ are less likely to be scattered into.

Now we consider the $f$ dependence of the two-magnon scattering rate. When $f$ is small, the two-magnon scattering rate increases as $f$ increases because more degenerate spin wave modes become available as $f$ increases. When $f$ further increases, the wavenumber $k$ of some degenerate spin wave modes exceeds $1/\xi$. This will decrease the overall two-magnon scattering rate because the degenerate spin wave modes with $k > 1/\xi$ are less likely to be scattered into, as discussed above. Furthermore, the portion of degenerate spin wave modes with $k > 1/\xi$ increases as $f$ continues to increase. When the impact of decreasing two-magnon scattering rate for degenerate spin wave modes with high $k$ surpasses the impact of increasing available degenerate spin wave modes, the overall two-magnon scattering rate will start to decrease as $f$ increases. Consequently, the nonlinear trend -- i.e., a ``hump'' -- in FMR linewidth $\Delta H_\mathrm{TMS}$ vs $f$ appears in Fig. \ref{fig:ip}.

However, the scenario discussed above can only happen when $\xi$ is large enough, because the wavenumber $k$ of degenerate spin wave modes saturates (i.e., reaches a limit) as $f$ approaches infinity. If the limit value of $k$ is smaller than $1/\xi$, the two-magnon scattering rate will increase monotonically as $f$ increases. In that case the hump feature will not appear. See our videos in the Supplemental Material that display the $f$ dependence of $\Lambda_{0k}$, $\delta_\alpha(\omega-\omega_k)$, $\frac{C_k(\xi)}{2\pi\xi^2}$, $\frac{\Lambda_{0k}C_k(\xi)\delta_\alpha(\omega-\omega_k)}{2\pi\xi^2}$, and $\Delta H_{\mathrm{TMS}}$ for various $\xi$ values.

Previous discussions of the hump feature are all based on the self-affine correlation function (Eq.~\ref{eq:selfaffine}). The main difference between the mounded correlation function (Eq.~\ref{eq:mounded}) and the self-affine correlation function (Eq.~\ref{eq:selfaffine}) is that the mounded correlation function has a peak when $\lambda$ is not much larger than $\xi$ as shown in Fig. \ref{fig:Ck}. This means when the wavenumber $k$ of degenerate spin wave modes enters (leaves) the peak region, two-magnon scattering rate will increase (decrease) much faster compared to the self-affine correlation function. In other words, the mounded correlation function can generate a narrower hump compared to the self-affine correlation function in the two-magnon linewidth $\Delta H_\mathrm{TMS}$ vs $f$ plot, which is shown in Fig. \ref{fig:ip_fit} (b, c).

\bibliographystyle{apsrev4-1}

\bibliography{refs.bib}

\end{document}